\documentstyle[12pt,a4wide]{article}
\newcommand{\beq}{ \begin{equation} }
\newcommand{\eeq}{ \end{equation} }
\newcommand{\nl}{ {\hfill \break} }
\newcommand{\np}{ {\newpage } }

\newcommand{\N}{ \mbox{\rm I$\!$N} }
\newcommand{\R}{ \mbox{\rm I$\!$R} }
\newcommand{\Diff}{ \mbox{\rm Diff} }
\newcommand{\lcm}{ \mbox{\rm lcm} }

\newcommand{\sign}{ \mbox{\rm sign} }

\newcommand{\artanh}{ \mbox{\rm artanh} }
\newcommand{\arcoth}{ \mbox{\rm arcoth} }

\newcommand{\MS}{ {\cal M\!S} }

\newcommand{\BlRZ }{1}
\newcommand{\Bl   }{2}
\newcommand{\Iv  }{{3}}
\newcommand{\bSch}{{4}}
\newcommand{\Sw  }{{5}}
\newcommand{\GMSS}{{6}}
\newcommand{\FM  }{{7}}
\newcommand{\Pa  }{{8}}
\newcommand{\aSch}{{9}}
\newcommand{\Ha }{{10}}
\newcommand{\Gu }{{11}}
\newcommand{\Ma }{{12}}
\newcommand{\Xa }{{13}}
\newcommand{\Ga }{{14}}
\newcommand{\Gi }{{15}}
\newcommand{\Ze }{{16}}
\newcommand{\aCh}{{17}}
\newcommand{\bCh}{{18}}
\newcommand{\aZh}{{19}}
\newcommand{\bZh}{{20}}

\begin{document}

\centerline{\Large \bf Conformal Coupling and Invariance}
\vspace{0.2truecm}
\centerline{\Large \bf in Different Dimensions}

\bigskip

\centerline{\bf {\large Martin Rainer}}

\vspace{0.5truecm}

\centerline{Projektgruppe Kosmologie, Mathematisches Institut}
\centerline{Universit\"at Potsdam, Am Neuen Palais 10}
\centerline{P.O.Box 601553, D-14415 Potsdam, Germany}

\vspace{0.15truecm}
\centerline{and}
\vspace{0.15truecm}

\centerline{Gravitationsprojekt, Universit\"at Potsdam}
\centerline{An der Sternwarte 16}
\centerline{D-14482 Potsdam, Germany}


\vspace{0.7truecm}

\begin{abstract}

\vspace{0.1truecm}
\hspace*{-0.75truecm}
Conformal transformations of the following kinds are compared:
(1) conformal coordinate transformations,
(2) conformal transformations of Lagrangian models
for a $D$-dimensional geometry, given by a Riemannian
manifold $M$ with metric $g$ of arbitrary signature,
and (3) conformal transformations of (mini-)superspace geometry.
For conformal invariance under this transformations
the following applications are given respectively:
(1) Natural time gauges for multidimensional geometry, (2)
conformally equivalent Lagrangian models for geometry coupled
to a spacially homogeneous scalar field, and (3) the conformal Laplace
operator on the  $n$-dimensional manifold $\cal M$ of
minisuperspace for multidimensional geometry and the
Wheeler de Witt equation.

The conformal coupling constant $\xi_c$ is critically distinguished
among arbitrary couplings $\xi$, for both,
the equivalence of Lagrangian models with $D$-dimensional
geometry and the conformal geometry on $n$-dimensional minisuperspace.

For dimension $D=3,4,6$ or $10$,
the critical number $\xi_c=\frac{D-2}{4(D-1)}$
is especially simple as a rational fraction.
\vspace*{0.7cm}
\nl

\end{abstract}

\np
\section{\bf Introduction}
\setcounter{equation}{0}
Recently conformal transformations between different
multidimensional geometrical models$^{\BlRZ-\Iv}$
and the question of conformal equivalence receive
increasing interest for cosmology. While cosmologist in recent time
apply conformal transformations already to minisuperspace and the
Wheeler de Witt (WdW) equation, several mathematical questions concerning
the structure of such transformations are, besides lacking
mathematical clarification of the (mini-)superspace construction,
still open, although Refs. \bSch\ and \Sw\ indicate some progress.
In Ref. \GMSS\ (and also Refs. therein)
conformal transformations have already
been discussed systematically within both,
the class of higher order gravity and that of
gravity coupled to a scalarfield, and between these two classes.
This paper is intended to contibute to an understanding
of conformal transformations of minisuperspace geometry as compared
to conformal transformations of ordinary geometry and coordinate
transformations. The conformal coupling constant $\xi_c$
will play a distinguished role for conformal invariance in different
contexts like equivalent Lagrangian models and the conformal Laplace
operator. Furthermore its dependence on the dimension of the underlying
space has interesting number theoretical properties,
distinguishing those low dimensions
which appear in prefered theories of the universe.

In Sec. 2 we compare
conformal transformations of the following kinds:
(1) conformal coordinate transformations,
(2) conformal transformations of Lagrangian models
for a $D$-dimensional geometry, given by a Riemannian
manifold $M$ with metric $g$ of arbitrary signature,
and (3) conformal transformations of (mini-)superspace geometry.

As application of invariance under (1), in Sec. 3 special emphasis is
put to compare natural time gauges for
multidimensional universes
given by the choices of
i) the synchronous time $t_s$ of the universe $M$,
ii) the conformal time $\eta_i$ of a universe with the only spacial factor
$M_i$,
iii) the mean conformal time $\eta$, given differentially as some
scale factor weighted average of $\eta_i$ over all $i$ and
iv) the harmonic time $t_h$, which will be used as specially
convenient in calculations on minisuperspace, since in this gauge
the minisuperspace lapse function is $N\equiv 1$.

As application of invariance under (2)
in Sec. 4 we examine conformal transformations between
conformally equivalent Lagrangian models for $D$-dimensional geometry
coupled to a spacially homogeneous scalar field.
The conformal coupling const $\xi_c$ here plays a distinguished
role. We consider as
example of special interest the conformal transformation between a
model with minimally coupled scalar field and an equivalent conformal model
with a conformally coupled scalar field, thus generalizing
previous results from Refs. \FM, \Pa\ and \aSch\ for $D=4$.

In Sec. 5 we derive the
unique conformal Laplace(-Beltrami)
operator on a (Pseudo) Riemannian manifold $\cal M$ of dimension $n$.
Though this had been given already  by construction
of a conformal WdW equation in Ref. \Ha,
and in the mathematical literature there is agreement
on a linear coupling $\Delta+aR$ of Laplacian $\Delta$, generalized
from the flat case, and Ricci curvature scalar $R$ on the underlying
manifold, there is sometimes some confusion$^\Gu$
about the proper choice of the coupling $a$
on an arbitrarily curved manifold.
Therefore here we prove that
$\Delta+aR$ is conformal if and only if $n>1$ and $a=-\xi_c$.

As application of equivariance w.r.t. tranformations (3)
in Sec. 6 we motivate a mini-superspace for multidimensional geometry
with minimally coupled scalar field,
and get
a first quantization of the energy constraint to the WdW equation
in a both generally covariant and conformally equivariant manner,
where the Laplace operator of Sec. 5 is essential ingredient.

Sec. 7 is devoted to a number theoretical examination of the  rational
number $\xi_c=\frac{D-2}{4(D-1)}$,
revealing distinguished dimensions $D=3,4,6$ and $10$.

Sec. 8 resumes the perspective of the present results.

\section{\bf Conformal Transformations}
\setcounter{equation}{0}
Generally we will have to distinguish between (1) conformal
{\em coordinate} transformations in $D$-dimensional geometry
(2) conformal transformations of $D$-dimensional {\em geometry},
especially in {\em Lagrangian models}, and
(3) conformal transformations of $n$-dimensional
{\em minisuperspace geometry}.
\vspace{0.3cm}

{\bf (1) Conformal transformation to new coordinates:}
\nl
We {\em fix the geometry} and transform the metric tensor components
conformally,
\begin{equation}
g_{i'j'}=e^{2f(x)}g_{kl},
\end{equation}
via a coordinate transform satisfying
\begin{equation}
dx'^{i}=e^{-f(x)}dx^i \quad{\rm or}\quad
\frac{\partial x'^{i}}{\partial x^j}=e^{-f(x)}\delta^i_j.
\end{equation}
Here the first fundamental form
\begin{equation}
ds^2=g_{i'j'}dx^{i'}dx^{j'}=g_{ij}dx^{i}dx^{j},
\end{equation}
and therefore the (inner) geometry, remains invariant,
though looking different in  different coordinate frames.
\vspace{0.3cm}

{\bf (2) Conformal transformations of ordinary ($D$-dimensional) geometry:}
\nl
Let us consider a differentiable manifold $M$.
On a (Pseudo-)Riemannian
geometry $g$ on $M$, conformal transformations will
be represented as Weyl transformations $g\mapsto e^{2f}g$ with
$f\in C^\infty (M)$.
For a structure on $M$
given by the metric components $g_{ij}$ and
(additional) scalar fields $(\phi^1,\ldots,\phi^k)$,
a second order {\em Lagrangian model} consists in a Lagrangian variation
principle
\begin{equation}
\delta S=0 \quad{\rm with}\quad S=\int_{M}  \sqrt{\vert g\vert}{L} d^D\!x
\end{equation}
with Lagrangian
\begin{equation}
L=L(g_{ij},\phi^1,\ldots,\phi^k;
    g_{ij,l},\phi^1_{,l},\ldots,\phi^k_{,l};
    g_{ij,lm}).
\end{equation}
Conformal transformation {\em of the Lagrangian model} keeps $M$ fixed as a
differentiable manifold,
but varies its additional structures conformally
\begin{equation}
(g_{ij},\phi^1,\ldots,\phi^k)\to (\hat g_{ij},\hat\phi^1,\ldots,\hat\phi^k),
\end{equation}
yielding a new variational principle by demanding
\begin{equation}
\sqrt{\vert g\vert}{L} \stackrel{!}{=} \sqrt{\vert \hat g\vert}{\hat L}
\end{equation}
for the new Lagrangian
\begin{equation}
\hat L=\hat L(\hat g_{ij},\hat\phi^1,\ldots,\hat\phi^k;
\hat g_{ij,l},\hat\phi^1_{,l},\ldots,\hat\phi^k_{,l};
\hat g_{ij,lm}).
\end{equation}
Therefore conformal transformations of
(Lagrangian models for) geometry (plus eventual scalar fields)
are performed in practice
on a {\em fixed coordinate patch} $x^i$ of $M$.
\vspace{0.3cm}

{\bf (3) Conformal transformation of minisuperspace geometry:}
\nl
In Ref. \Sw\ superspace is defined as the geometries
$\mbox{Met}(M)/\Diff(M)$. We will not consider the
here the difficult question of $\Diff(M)$-equivalence.
Let us assume that this is solved by
a kind of general coordinate invariance of superspace.
In Ref. \bSch\ it was shown how the space of Riemannian metrics
$\mbox{Met}(M)$ can be equipped with a metric $G$.
This is what we will do here.
Let us pick some superspace coordinates $\chi^A$ indexed by $A$ within
an appropriate index set.
Then we consider
\begin{equation}
G=G_{AB}d\chi^A\otimes d\chi^B
\end{equation}
as defined via
\begin{equation}
G_{AB}:=G_{ijkl}h^{ij}_A h^{kl}_B,
\end{equation}
where
\begin{equation}
G_{ijkl}:=g_{ik}g_{jl}+g_{il}g_{jk}-g_{ij}g_{kl}.
\end{equation}
Note the similarity of Eq. (2.11) to the usual first Christoffel
symbols.
Both $G_{ijkl}$ and $h^{ij}_A$ are covariant 4- resp. 2-tensor components
with respect to usual coordinate transforms. Therefore $G_{AB}$
is independent of coordinates on $M$.

The components $h^{ij}_A$ define a generalized soldering form
$\theta:=h^{ij}_A  e_{ij}\otimes d\chi^A$, where
2-tensors $e_{ij}=h^A_{ij}\frac{\partial}{\partial \chi^A}$
are given by  components $h^A_{ij}$ dual to $h^{ij}_A$.

Eq. (2.11) singles out a special class of supermetrics,
and, together with general superspace covariance, yields
a reduced superspace.
However without further restriction
its index set would be still infinite.

A further reduction of superspace to yielding finite dimensions is
is well defined for
a class of metrics of multidimensional type.
Here a geometry is described on a (Pseudo-) Riemannian manifold
$$
M=\R\times M_1 \times\ldots\times M_n,  \qquad D:=\dim M=1+d_1+\ldots+d_n,
$$
\begin{equation}
g\equiv ds^2 = -e^{2\gamma} dt\otimes dt
     + \sum_{i=1}^{n} a_i^2 \, ds_i^2,
\end{equation}
where    $ a_i=e^{\beta^i} $
is the scale factor of the $d_i$-dimensional space $M_i$
with the first fundamental form
$$
ds_i^2
=g^{(i)}_{{k}{l}}\,dx^{k}_{(i)} \otimes dx^{l}_{(i)}.
$$
Then the scalefactors $e^{\beta^A}$, $A=1,\ldots,n$,
yield (reduced) supercoordinates
\beq
\chi^A:=e^{2\beta^A},\qquad A=1,\ldots,n.
\eeq
The {\em minisuperspace} ${\cal M\!S}(M)$ over $M$
is then defined by  minisuperspace coordinates $\beta^1,\ldots,\beta^n$
subject to the principle of general covariance w.r.t. minisuperspace
coordinate transformations.

Invariance of (2.10)
under conformal transformations (2) with $g\mapsto e^{2f}g$
yields invariance under
\begin{equation}
h^{ij}_A\mapsto e^{-2f}h^{ij}_A,
\end{equation}
which corresponds to invariance
under supercoordinate transformations
\begin{equation}
\chi^A\mapsto e^{2f}\chi^A.
\end{equation}
The conformal weight differs from that of an analogous ordinary
coordinate transformation (2.2) by a factor $-2$,
corresponding to the fact that
$h^{ij}_A$ contragrediently relates supervectors to 2-tensors.

For a minisuperspace ${\cal M}=\MS(M)$ from (2.13)
the supercoordinate tranformations (2.15)
correspond via (2.13) to translations
of the minisuperspace coordinates
\beq
\beta^i\to\beta^i+f.
\eeq
So conformal transformations (2) of multidimensional geometry $M$
yield just supercoordinate transformations in $\cal M$.

Well distinguished from the latter,
conformal transformations {\em of the minisuperspace geometry}
\begin{equation}
G = G_{ij}d\beta^i\otimes d\beta^j
\end{equation}
are given by
\beq
G\mapsto {}^{f}\!G:=e^{2f}G
\eeq
with $f\in C^\infty (\cal M)$.
\vspace{0.3cm}

So far we demonstrated the necessity to distinguish transformations
(3) against (2), in analogy to the difference between
transformations (2) and (1).

Applications of invariance under transformations (2) and (3) will
be given later. A special application of transformations (1) are time gauge
transformations, from arbitrarily given coordinates to
one of the natural time gauges considered in the following section.

\np

\section{\bf Natural Times in Multidimensional Geometry}
\setcounter{equation}{0}
Let us consider a multidimensional geometry
like in Eq. (2.12) and
compare different choices of time $t$  in Eq. (2.12).
The time gauge is determined by the function $\gamma$.
There exist few natural time gauges from the physical point of view.

i) The {\em synchronous time gauge}
\begin{equation}
\gamma\equiv 0,
\end{equation}
for which $t$ in Eq. (2.12) is the proper time $t_s$ of the universe.
The clocks of geodesically comoved observers go synchronous to that
time.

ii) The {\em conformal time gauges} on $\R\times M_i\subset M$
\begin{equation}
\gamma\equiv \beta^i,
\end{equation}
for which $t$ in Eq. (2.12) is the conformal time $\eta_i$ of $M_i$
for some $i\in \{1,\ldots,n\}$,
given by
\begin{equation}
d\eta_i=e^{-\beta^i}dt_s.
\end{equation}

iii) The {\em mean conformal time gauge} on $M$:
\nl
For $n>1$ and $\beta^2\neq\beta^1$
on $M$ the usual concept of a conformal time does no longer apply.
Looking for a generalized ``conformal time" $\eta$ on $M$, we set
\begin{equation}
d:=D-1=\sum_{i=1}^n d_i
\end{equation}
and consider the gauge
\begin{equation}
\gamma\equiv \frac{1}{d}\sum_{i=1}^n d_i\beta^i,
\end{equation}
which yields a time $t\equiv\eta$ given by
\begin{equation}
d\eta=\left( \prod_{i=1}^n a_i^{d_i} \right)^{-1/d} dt_s.
\end{equation}
Here $\prod_{i=1}^n a_i^{d_i}$ is proportional to the volume of
$d$-dimensional spacial sections in $M$
and the relative time scale factor
\begin{equation}
\left( \prod_{i=1}^n a_i^{d_i} \right)^{1/d}
=e^{\frac{1}{d}\sum_{i}d_i\beta^i}
\end{equation}
is given by a scale exponent, which is the dimensionally weighted arithmetic
mean of the spacial scaling exponents of spaces $M_i$.
It is
\begin{equation}
(dt_s)^d= e^{\sum_{i}d_i\beta^i} d\eta^d.
\end{equation}
Since on the other hand by Eq. (3.3) we have
\begin{equation}
(dt_s)^d=\otimes_{i=1}^n \left( e^{\beta^i}d\eta_i \right)^{d_i},
\end{equation}
together with Eq. (3.8) we yield
\begin{equation}
(d\eta)^d=e^{-\sum_{i}d_i\beta^i}
\otimes_{i=1}^n \left( e^{\beta^i}d\eta_i \right)^{d_i}.
\end{equation}
So the time $\eta$ is a mean conformal time, given differentially
as a dimensionally scale factor weighted geometrical tensor
average of the conformal times $\eta_i$.
An alternative to the mean conformal time $\eta$
is given by a similar differential averaging like Eq. (3.10), but weighted
by an additional factor of $e^{(1-d)\sum_{i}d_i\beta^i}$.
This is gauge is described in the following.

iv) The {\em harmonic time gauge}
\begin{equation}
\gamma\equiv\gamma_h:=\sum_{i=1}^n d_i\beta^i
\end{equation}
yields the time $t\equiv t_h$, given by
\begin{equation}
dt_h=\left( \prod_{i=1}^n a_i^{d_i} \right)^{-1} dt_s
=\left( \prod_{i=1}^n a_i^{d_i} \right)^{\frac{1-d}{d}} d\eta.
\end{equation}
In this gauge any function $\varphi$ with $\varphi (t,y)=t$
is harmonic, i.e. $\Delta[g] \varphi =0$, and
the minisuperspace lapse function is $N\equiv 1$.
The latter is especially convenient when we work in
minisuperspace.

\section{\bf Conformally Equivalent Lagrangian Models}
\setcounter{equation}{0}

Now we want to study the effect of transformations (2) in more detail.
One application of special interest is the transformation from
a Lagrangian model with minimally coupled scalarfield to a conformally
equivalent one with non-minimally coupled scalarfield and vice versa.

Let us follow Ref. \Ma\  and consider an action of the kind
\begin{equation}
S=\int d^Dx\sqrt{\vert g\vert}(F(\phi,R)-\frac{\epsilon}{2}(\nabla\phi)^2).
\end{equation}
With
\begin{equation}
\omega:=\frac{1}{D-2}\ln(2\kappa^2
			  \vert\frac{\partial F}{\partial R}\vert)+C
\end{equation}
the conformal factor
\begin{equation}
e^{\omega}=
[2\kappa^2\vert\frac{\partial F}{\partial R}\vert]^\frac{1}{D-2}e^{C}
\end{equation}
yields a conformal transformation from $g_{\mu\nu}$
to the minimally coupled metric
\begin{equation}
\hat g_{\mu\nu}=e^{2\omega}g_{\mu\nu}.
\end{equation}

Especially let us consider in the following actions, which are linear
in $R$. With
\begin{equation}
F(\phi,R)=f(\phi)R-V(\phi).
\end{equation}
the action is
\begin{equation}
S=\int d^Dx\sqrt{\vert g\vert}(f(\phi)R-V(\phi)
-\frac{\epsilon}{2}(\nabla\phi)^2).
\end{equation}

The minimal coupling metric
is then related to the conformal one by (4.4) with
\begin{equation}
\omega=\frac{1}{D-2}\ln(2\kappa^2
\vert f(\phi)\vert)+C
\end{equation}
The scalar field in the minimal coupling model is
$$
\Phi=\kappa^{-1}\int d\phi\{\frac{\epsilon(D-2)f(\phi)+2(D-1)(f'(\phi))^2}
			       {2(D-2)f^2(\phi)}\}^{1/2}  =
$$
\begin{equation}
=(2\kappa)^{-1}\int d\phi\{\frac{2\epsilon f(\phi)+\xi_c^{-1}(f'(\phi))^2}
			       {f^2(\phi)}\}^{1/2},
\end{equation}
where
\begin{equation}
\xi_c:=\frac{D-2}{4(D-1)}
\end{equation}
is the conformal coupling constant.

For the following we define $\sign x$ to be $\pm 1$ for $x\geq 0$ resp.
$x<0$.
Then with the new minimal coupling potential
\begin{equation}
U(\Phi)=(\sign f(\phi))\ [2\kappa^2\vert f(\phi)\vert]^{-D/D-2}V(\phi)
\end{equation}
the corresponding minimal coupling action is
\begin{equation}
S=\sign f\int d^Dx\sqrt{\vert \hat{g}\vert}\left(-\frac{1}{2}
	  [(\hat{\nabla}\Phi)^2-\frac{1}{\kappa^2}\hat{R}]-U(\Phi)\right).
\end{equation}

Example 1:
\begin{equation}
f(\phi)=\frac{1}{2}\xi\phi^2,
\end{equation}
\begin{equation}
V(\phi)=-\lambda\phi^\frac{2D}{D-2}.
\end{equation}
Substituting this into Eq. (4.10) the corresponding
minimal coupling potential
$U$ is constant,
\begin{equation}
U(\Phi)=(\sign \xi)\ \vert\xi\kappa^2\vert^{-D/D-2} \,\lambda.
\end{equation}
It becomes zero precisely for $\lambda=0$, i.e. when $V$ is zero.
With
\begin{equation}
f'(\phi)=\xi\phi
\end{equation}
we obtain
$$
\Phi=\kappa^{-1}\int d\phi
\left\{
\frac{ ({\epsilon\over\xi} + {1\over{\xi_c}} ) \phi^2}{\phi^4}
\right\}^\frac{1}{2}
=\left(\kappa\sqrt\xi\right)^{-1}
\sqrt{\frac{1}{\xi_c}+ {\epsilon\over\xi} }
\int d\phi\frac{1}{\vert\phi\vert}
$$
\begin{equation}
=\kappa^{-1}\sqrt{\frac{1}{\xi_c}+{\epsilon\over\xi} }\,\ln\vert\phi\vert
+C
\end{equation}
for $-\frac{\xi}{\epsilon}\geq\xi_c$.
Note that for
\begin{equation}
\frac{\xi}{\epsilon}=-\xi_c,
\end{equation}
e.g. for $\epsilon=-1$ and conformal coupling, we have
\begin{equation}
\Phi=C.
\end{equation}
Thus here the conformal coupling theory is equivalent to
a theory without scalarfield.

For $-\frac{\xi}{\epsilon}<\xi_c$ the field $\Phi$ would become
complex and, for imaginary $C$, purely imaginary.

In any case the integration constant $C$ may be a function of the
coupling $\xi$ and the dimension $D$.

Example 2:\nl
\begin{equation}
f(\phi)=\frac{1}{2}(1-\xi\phi^2),
\end{equation}
\begin{equation}
V(\phi)=\Lambda.
\end{equation}
Then the constant potential $V$ has its minimal coupling correspondence in a
non constant $U$, given by
\begin{equation}
U(\Phi)=\pm \Lambda \vert\kappa^2 (1-\xi\phi^2) \vert^{-D/D-2}
\end{equation}
respectively
for $\phi^2<\xi^{-1}$ or $\phi^2>\xi^{-1}$.

Let us set in the following
\begin{equation}
\epsilon=1.
\end{equation}
Then with
\begin{equation}
f'(\phi)=-\xi\phi
\end{equation}
we obtain
\begin{equation}
\Phi=\kappa^{-1}\int d\phi\{\frac{1+c\,\xi\phi^2}
			       {(1-\xi\phi^2)^2}\}^{1/2},
\end{equation}
where
\begin{equation}
c:=\frac{\xi}{\xi_c}-1.
\end{equation}

For $\xi=0$ it is $\Phi=\kappa^{-1}\phi +A$, i.e. the coupling remains
minimal.

To solved this integral for $\xi\neq 0$, we substitute $u:=\xi\phi^2$.

To assure a solution of (4.24) to be real, let us assume $\xi\geq\xi_c$
which yields $c\geq 0$.

Then we obtain
$$
\Phi=\frac{\sign(\phi)}{2\kappa\sqrt{\xi}}
		  \int{\frac {\sqrt {u^{-1}+c}}{\vert1-u\vert}}du
+C_{<\atop>}
$$
$$
=\frac{\sign((1-u)\phi)}{2\kappa\sqrt{\xi}}
[-\sqrt {c}\ln (2\,\sqrt {c}\sqrt {1+cu}\sqrt {u}+2\,cu+1)+
$$
$$
\sqrt {1+c}
\ln ({\frac{2\,\sqrt{1+c}\sqrt{1+cu}\sqrt {u}+2\,cu+1+u}{\vert1-u\vert}})]
+C_{<\atop>}
$$
$$
=\frac{\sign((1-\xi\phi^2)\phi)}{2\kappa\sqrt{\xi}}
\{-\sqrt {c}\ln (2\,\sqrt {c}\sqrt {1+c\xi\,\phi^{2}}
\sqrt {\xi} \vert\phi\vert
+2\,c\xi\,\phi^{2}+1)
$$
$$
+ \sqrt {1+c}\ln ({\frac {2\,\sqrt {1+c}
\sqrt{1+c\xi\,\phi^{2}} \sqrt {\xi} \vert\phi\vert
+2\,c\xi\,\phi^{2}+1+\xi\,\phi^{2}}
{\vert 1-\xi\,\phi^{2}\vert}})\}
+C_{<\atop>}
$$
$$
=\frac{\sign((1-\xi\phi^2)\phi)}{2\kappa\sqrt{\xi}}
\ln
\frac{ [2\,\sqrt {1+c} \sqrt{1+c\xi\,\phi^{2}}\sqrt {\xi}\vert\phi\vert
	    +(2\,c+1)\xi\,\phi^{2}+1]^{\sqrt {1+c}}  }
{ [2\,\sqrt {c}\sqrt {1+c\xi\,\phi^{2}}\sqrt {\xi}\vert\phi\vert
	     +2\,c\xi\,\phi^{2}+1]^{\sqrt{c}}
\cdot {\vert 1-\xi\,\phi^{2}\vert}^{\sqrt {1+c}}  }
$$
\begin{equation}
+C_{<\atop>}.
\end{equation}
The integration constants $C_{<\atop>}$
for $\phi^2<\xi^{-1}$ and $\phi^2>\xi^{-1}$ respectively
may be arbitrary functions of $\xi$ and the dimension $D$.

The singularities of the transform $\phi\to\Phi$ are located at
$\phi^2=\xi^{-1}$.
An expression corresponding to (4.26) with $D=4$ qualitatively has
already been given in Ref. \FM.

If the coupling is conformal $\xi=\xi_c$,
i.e. $c=0$, the expressions (4.26) simplify to
\begin{equation}
\kappa\Phi=\frac{1}{\sqrt{\xi_c}} [(\artanh\sqrt{\xi_c}\phi)+c_<]
\end{equation}
for $\phi^2<\xi^{-1}_c$ and to
\begin{equation}
\kappa\Phi=\frac{1}{\sqrt{\xi_c}} [(\arcoth\sqrt{\xi_c}\phi)+c_>]
\end{equation}
for $\phi^2>\xi^{-1}_c$.

Then the inverse formulas expressing the conformal field $\phi$ in terms of
the minimal coupling field $\Phi$ are
\begin{equation}
\phi=\frac{1}{\sqrt{\xi_c}} \left[ \tanh(\sqrt{\xi_c}\kappa\Phi-c_<) \right]
\end{equation}
with $\phi^2<\xi^{-1}_c$ and
\begin{equation}
\phi=\frac{1}{\sqrt{\xi_c}} \left[ (\coth(\sqrt{\xi_c}\kappa\Phi-c_>) \right]
\end{equation}
with $\phi^2>\xi^{-1}_c$ respectively.
This result agrees with Ref. \Xa.
For $D=4$ it has been obtained earlier in Refs. \Pa, \aSch\ and \Ga.
In Ref. \Ga\ it has been shown for $D=4$,  that while the minimal
coupling model shows a curvature singularity, the conformal coupling
model with $\phi$ of Eq. (4.29) has no such singularity.

The conformal factor is according to Eqs. (4.7) and (4.19) given by
\begin{equation}
\omega=\frac{1}{D-2}\ln(\kappa^2 \vert 1-\xi_c\phi^2 \vert)+C.
\end{equation}

The singularity
of the conformal transformation (4.31)
at $\phi^2=\xi^{-1}_c$ separates different regions in $\phi$ where
conformal equivalence
between the minimal and conformal coupling model holds.
Eqs. (4.29) and (4.30)  illustrate the qualitatively
different behavior in the two regions.
In Ref. \BlRZ\ this qualitative difference have also been found
in multidimensional solutions of the respective models.

Note finally that, if e.g. time is harmonic in the minimal coupling model
\begin{equation}
\tau\equiv t^{(m)}_h,
\end{equation}
in the conformal model it cannot be expected to be harmonic either,
i.e. in general
\begin{equation}
\tau\neq t^{(c)}_h.
\end{equation}
Natural time gauges are not preserved by conformal transformations (2).
Usually they have to be calculated by a coordinate transformation
in each of the equivalent models separately.

\section{\bf The Conformal Laplace Operator}
\setcounter{equation}{0}
In this section we search for a
linear combination
\begin{equation}
\Delta_a = \Delta  + aR
\end{equation}
of the Laplace-Beltrami operator  $\Delta=\Delta [G]$ and
the Ricci scalar curvature $R=R[G]$ of an $n$-dimensional manifold
$\cal M$, such that
$\Delta_a$ is not only a
generally covariant but also a conformal operator of weight $-2$, which
furthermore transforms according to the conjugate representation $D_b$ of
Weyl transformations of weight $b$ on the Hilbertspace $\cal H(M)$.
With $f\in C^\infty(M)$, the latter transform
$\cal H:=\cal H(M)$ to ${\cal H}^{f}:=e^{bf}\cal H$.
Then the conformal operator on ${\cal H}^f$ is
\begin{equation}
\Delta^f_a = e^{(b-2)f}\Delta_a e^{-bf}
\end{equation}
where $\Delta^f_a = \Delta^f  + aR^f$ with
\begin{equation}
\Delta^f=^f\!G^{ij}\nabla^f_i\nabla^f_j.
\end{equation}
Here the covariant derivative $\nabla^f$ is determined by the connection
$\Gamma^f$ w.r.t. the metric $G^f$.
Since the components of the inverse metric are
\begin{equation}
^f\!G^{ij}=e^{-2f}G^{ij},
\end{equation}
the connection coefficients are
$$
^f\!\Gamma^k_{ij}={\frac{1}{2}}{}^f\!G^{kl}
\left\{
^f\!G_{li,j}+^f\!G_{li,j}-^f\!G_{ij,l}
\right\}
$$
\begin{equation}
=\Gamma^{k}_{ij}+
\left\{
\delta^k_i f_{,j}+\delta^k_j f_{,i} -G_{ij} f^{,k}
\right\}
\end{equation}
and the Ricci scalar w.r.t. $^f\!G$ is
$$
^f\! R
= e^{-2f} G^{cd}[R_{cd}-2(n-1) f_{,cd}-(n-1)(n-2)f_{,c}f_{,d}
+2(n-1)\Gamma^e_{cd} f_{,e}],
$$
\begin{equation}
= e^{-2f} \left\{
R-2(n-1)\Delta f-(n-1)(n-2)f^{,k}f_{,k} \right\}.
\end{equation}

On ${\cal H}^{f}$ we find
$$
\Delta^f= ^f\!G^{ij}\ ^f\!\nabla_i\ \partial_j
$$
\begin{equation}
=e^{-2f}\Delta-e^{-2f}G^{cd} \left\{
f_{,i}\partial_j+f_{,j}\partial_i-\Gamma_{ij}f^k\partial_{k} \right\}
\end{equation}
in terms of the original metric $G$ and its Laplacian $\Delta$ on
acting on ${\cal H}$.
Thus we obtain
\begin{equation}
\Delta^f\Psi^f=e^{(b-2)f}
\left\{
\Delta\Psi+[2(b-1)+n]f^{,k}\Psi_{,k}
+[b\Delta f+(b+n-2)bf^kf_k]\Psi
\right\}
\end{equation}
and together with Eq. (5.6) it is
\begin{equation}
\Delta^f_a\Psi^f=e^{(b-2)f}
\left\{
\Delta_a\Psi+[2(b-1)+n]f^{,k}\Psi_{,k}
+[A \Delta f+ B f^k f_k]\Psi
\right\}
\end{equation}
with coefficients
\begin{equation}
A = b-2(n-1)a\ \mbox{and}\ B = (b+n-2)b-(n-1)(n-2)a.
\end{equation}
Vanishing of the $f^{,k}\Psi_{,k}$ term in Eq. (5.9) requires
\begin{equation}
b=1-\frac{n}{2}=-\frac{n-2}{2},
\end{equation}
which then yields the coefficients
\begin{equation}
A = -\frac{1}{2}\{4(n-1)a+(n-2)\}
\end{equation}
and
\begin{equation}
B = -\frac{n-2}{4}\{4(n-1)a+(n-2)\}
\end{equation}
both proportional to $4(n-1)a+(n-2)$.
Then for $n\neq 1$ their vanishing requires
\begin{equation}
a=-\frac{n-2}{4(n-1)}\equiv -\xi_c.
\end{equation}
For $n=1$ the condition $(5.11)$ implies $b=\frac{1}{2}$ where
$A\neq 0 \neq B$, while vanishing of $A$ and $B$ according to Eq. (5.10)
implies $b=0$ where condition (5.11) is violated. Thus for $n=1$
there is no conformal operator (5.1) for any value of $a$.
This is because every $1$-dimensional manifold is intrinsically flat and
hence $R\equiv 0$, while the representation $D_b$ is non trivial for
$b=\frac{1}{2}$.

Note that condition (5.11) excludes the trivial representation $b=0$
in all dimensions except $n=2$, where $a=b=0$ and
$\Psi^f =\Psi$ for all $f$,
since any $2$-dimensional manifold is conformally flat.

We find that  the operator (5.1) is conformally invariant
if and only if $n\neq 1$ and the values of $a$ and $b$ are given by
Eq. (5.11) and (5.14).

\section{\bf Minisuperspace and the WdW Equation}
\setcounter{equation}{0}
If we assume within a multidimensional geometry (2.12) that
$M_i$ are Einstein spaces,
they satisfy the equations
\begin{equation}
R^{(i)}_{kl}=\lambda_i g^{(i)}_{kl},
\end{equation}
and hence
\begin{equation}
R^{(i)}=\lambda_i d_i.
\end{equation}
Here the Ricci tensor and scalar are defined as usual by
\begin{equation}
R_{\mu\nu}:=R^\lambda_{\mu\lambda\nu}
\quad{\rm and}\quad
R:=R^\mu_\mu.
\end{equation}
If furthermore $M_i$ is of constant curvature, then
\begin{equation}
ds_i^2
=\frac{1}{(1+\frac{1}{4}K_i r_i^2)^2}\sum_{k=1}^{d_i}
dx^{k}_{(i)} \otimes dx^{k}_{(i)},
\end{equation}
with radial variable
$r_i=\sqrt{\sum_{k=1}^{d_i}\left(x^k_{(i)}\right)^2}$
and constant
sectional curvature, normalized with $K_i=\pm 1$ for positive and negative
$K_i$ respectively. In the flat case $K_i=0$.
Then the Riemann tensor of $M_i$ is
\begin{equation}
R^{(i)}_{klmn}=K_i(g^{(i)}_{km}g^{(i)}_{ln}-g^{(i)}_{kn}g^{(i)}_{lm}).
\end{equation}
Ricci tensor and scalar are then given by Eq. (6.1) and (6.2) with
\begin{equation}
\lambda_i\equiv K_i (d_i-1).
\end{equation}
For the geometry (2.12) the Ricci scalar curvature of $M$ is
\begin{equation}
R=e^{-2\gamma}\left \{
\left [ \sum_{i=1}^{n} (d_i \dot\beta^i) \right ]^2
+ \sum_{i=1}^{n}
d_i[ {(\dot\beta^i)^2-2\dot\gamma\dot\beta^i+2\ddot \beta^i} ]
\right \}
+\sum_{i=1}^{n} R^{(i)} e^{-2\beta^i}.
\end{equation}

Let us now consider a variation principle with the action
\begin{equation}
S=S_{EH}+S_{GH}+S_{M},
\end{equation}
where
\[
S_{EH}=\frac{1}{2\kappa^2}\int_{M}\sqrt{\vert g\vert} R\, dx
\]
is the Einstein-Hilbert action,
\[
S_{GH}=\frac{1}{\kappa^2}\int_{\partial M}\sqrt{\vert h\vert} K \, dy
\]
is the Gibbons-Hawking boundary term$^\Gi$, with $K$ the trace
of the second fundamental form according to the ADM decomposition
(this term is required for canceling second time derivatives
in the equations of motion), and $S_M$ some matter term.

Let us consider here a matter term $S_M$ corresponding to
a minimally coupled
scalar field $\Phi$ with potential $U(\Phi)$.
Then the
variational principle of (6.8) is  equivalent to
a Lagrangian variational principle
over the minisuperspace $\cal M$ and the scalar field $\Phi$,
$$
S=\int Ldt,
$$
$$
L=\frac{1}{2}{\mu}
e^{-\gamma+\sum_{i=1}^{n}d_i\beta^i}
\left \{
\sum_{i=1}^{n}{d_i(\dot\beta^i)^2}
-[\sum_{i=1}^{n}{d_i\dot\beta^i}]^2
+\kappa^2{\dot \Phi}^2
\right \}
$$
\begin{equation}
+\frac{1}{2}{\mu}
e^{\gamma+\sum_{i=1}^{n}d_i\beta^i}R^{(i)}e^{-2\beta^1}
-{\mu}\kappa^2 e^{\gamma+\sum_{i=1}^{n}d_i\beta^i}
U(\Phi),
\end{equation}
where
\begin{equation}
\mu:=\kappa^{-2}\prod_{i=1}^{n}\sqrt{\vert\det g^{(i)}\vert}.
\end{equation}
It is a convenient proceedure of cosmologists,
to extend the minisuperspace $\cal M$ of pure geometry directly by an
additional dimension from the scalar field $\Phi$ as further
minisuperspace coordinate, yielding an enlarged
minisuperspace $\MS:=\MS(M,\Phi)$.

Let us define a metric on $\MS$, given in
coordinates $\beta^i$, $i=1,\ldots,n+1$ with $\beta^{n+1}:=\kappa\Phi$.
We set
\begin{equation}
G_{n+1\,i}=G_{i\,n+1}:=\delta_{i\,n+1}
\ \mbox{and}\
G_{kl}:=d_k \delta_{kl}-d_k d_l
\end{equation}
for $i=1,\ldots,n+1$ and $k,l=1,\ldots,n$,
thus defining the components
$G_{ij}$ of the minisuperspace metric
\begin{equation}
G = G_{ij}d\beta^i\otimes d\beta^j.
\end{equation}
Note that the signature of $\cal M$ is Lorentzian for $n>1$,
and $G_{11}<0$ for $d_1>1$ implies that the signature of $\MS$ is
Lorentzian not only for $n>1$ but also for $n=1$ if $d_1>1$.
If there is at least one
(e.g. compact ''internal'') extra factor space, i.e. $n>1$, then $\cal M$
has Lorentzian signature $(-,+,\ldots,+)$.

\np
After diagonalization
of (6.11) by a minisuperspace coordinate transformation
$\beta^i\to\alpha^i$ ($i=1,\ldots,n$), there is just one new
coordinate, say $\alpha^1$, which corresponds to the unique
negative eigenvalue of $G$. With a further (sign preserving)
coordinate rescaling, $G$ is equivalent to the
Minkovsky metric. Hence $\cal M$ is conformally flat.

While $\beta^i\to\alpha^i$ is only a coordinate transformation
(1) on $\cal M$ or $\MS$, it transforms a multidimensional geometry
(2.12) with scale exponents $\beta^i$ to another geometry
of the same multidimensional type (2.12),
i.e. with the same $d_i$ and $ds^2_i$,
but new scale exponents $\alpha^i$ of the factor spaces
$M_i$. We can always perform the diagionalization of (6.11) such
that $\alpha^1$ and hence $M_1$ belongs to the unique
negative eigenvalue of $G$. This $M_1$ is identified as ''external''
space. The scale factors of the ''internal''
spaces $M_2,\ldots,M_n$ and $\Phi$ contribute only
positive eigenvalues of $\cal M$ reps. $\MS$.
(For $n=1$ there are no ''internal'' spaces, but $G_{11}<0$ for
$d_1>1$ still provides a negative eigenvalue that is distinguished
at least against the additional positive eigenvalue from
$\Phi$ in $\MS$.)
$\alpha^1$ assumes in $\cal M$ or $\MS$ the role played by time
in usual geometry and quantum mechanics.
In this way ''external'' space
is distinguished against the ''internal'' spaces, because its scale
factor provides a natural ''time'' coordinate on $\cal M$.
If in the multidimensional geometry (2.12) $M_1$ with $\alpha^1$ is
strictly expanding
w.r.t. time $t$, then the
''minisuperspace time'' $\alpha^1$ can be considered in the geometry $g$
as a time equivalent to $t$. So the Lorentzian structure of $\cal M$
finally provides with expanding $M_1$ a natural ''arrow of time''$^\Ze$.

Now we define
a minisuperspace lapse function by
\begin{equation}
N:=e^{\gamma-\sum_{i=1}^{n}d_i\beta^i}
\end{equation}
and a minisuperspace potential $V=V(\beta^i)$ via
\begin{equation}
V:=-\frac{\mu}{2}
\left( \sum_{i=1}^{n} R^{(i)} - 2\kappa^2 U(\Phi) \right)
e^{-2\beta^i+\gamma+\sum_{j=1}^{n}d_j\beta^j}.
\end{equation}
Then
\begin{equation}
L=N\{\frac{\mu}{2}N^{-2}G_{ij}\dot\beta^i\dot\beta^j-V\}.
\end{equation}
Here $\mu$ is the mass of a classical particle in minisuperspace.
Note that $\mu^2$ is proportional to the volumes of spaces $M_i$,
which is a purely geometrical datum on $M$ but not on $\cal M$ or
$\MS$.

In the harmonic time gauge,
the equations of motion from Eq. (6.15) are given by
\begin{equation}
\mu G_{ij}\ddot\beta^j=-\frac{\partial V}{\partial\beta^i}
\end{equation}
with the energy constraint
\begin{equation}
\frac{\mu}{2}G_{ij}\dot\beta^i\dot\beta^j+V=0.
\end{equation}

In the following we should forget,
whether some of the coordinates $\beta^i$ are of nongeometric
origin, since all will be treated equally. Therefore we will understand
by $\cal M$ some minisuperspace (no matter whether it it has actually
been constructed as some $\MS$ or not).

Canonical quantization has been considered e.g.
in Refs. \Ha, \aCh\ and \bCh. It essentially consists in
replacing the constraint equation (6.17) by
the WdW equation
\begin{equation}
\left(-\frac{1}{2}\left[\Delta+aR\right]+V\right)\Psi = 0
\end{equation}
where $\Psi$ is a wave function from a distribution space ${\cal S}^\ast$,
which is the dual of the test function space ${\cal S}\subset{\cal H}$,
dense in the Hilbert space ${\cal H}={\cal H(M)}$.
(Often one might think of ${\cal S}$ as the Schwartz space and
correspondingly of
${\cal S}^\ast$ as the space of tempered distributions over ${\cal S}$.
Note however that the proper choice of ${\cal S}$ depends on the Hamiltonian
$H$, and more specifically on the shape of the potential $V$.)

The Lagrangian (6.15) is invariant under
arbitrary time reparametrization $h\in \mbox{Diff}\,{\R}$
acting via
\[
h(\beta^i)(t):=\beta^i(h(t))\quad \mbox{and}\quad h({N})(t)
:={N}(h(t))\frac{dh}{dt}
\]
on minisuperspace coordinates $\beta^i$ and the lapse function $N$.

We set in the following
\begin{equation}
{N} =: e^{-2f}
\end{equation}
and admit $f\in C^\infty(\cal M)$ to be an arbitrary smooth function on
$\cal M$.

In the time gauge given by $f$
the Lagrangian is
\begin{equation}
L^f:= \frac{\mu}{2} {^{f}}\!G_{ij}(\beta)\dot{\beta}^i\dot{\beta}^j
- V^f(\beta )
\end{equation}
and the energy constraint is
\begin{equation}
E^f:= \frac{\mu}{2} {^{f}}\!G_{ij}(\beta )\dot{\beta}^i\dot{\beta}^j
+ V^f(\beta ) = 0,
\end{equation}
where
\[
^{f}\!G = e^{2f}G \ \mbox{and} \ V^f = e^{-2f}V.
\]
With canonical momenta
\begin{equation}
\pi_{i} = \frac{\partial L^f}{\partial\dot{\beta}^{i}} =
\mu G_{ij}^{f}\dot\beta^{j}
\end{equation}
this is equivalent to the Hamiltonian system given by
\begin{equation}
H^f = \frac{1}{2\mu}(^f\!G)^{ij}\pi_i\pi_j + V^f
\end{equation}
and the energy constraint
\begin{equation}
H^f = 0.
\end{equation}
There the inverse of the minisuperspace metric is given by
$^{f}\!G^{-1} = e^{-2f} G^{-1}$, where for the system with Eq. (6.11)
the components of $G^{-1}$ are
\begin{equation}
G^{ij} = \frac{\delta_{ij}}{d_i} + \frac{1}{1-\sum_{i=1}^{n}d_i}.
\end{equation}
At quantum level $H^f$ has to be replaced by an operator
$\hat{H}^f$, acting by the energy constraint
\begin{equation}
\hat{H}^f\Psi^f = 0
\end{equation}
on  $\Psi^f\in {\cal S}^{\ast f}$, where ${\cal S}^{\ast f}$ is given by
the action of a represention $D_b$ of
$C^\infty(\cal M)$ with conformal weight $b$ on
$\Psi\in{{\cal S}^{\ast}}$, i.e. for $f\in C^\infty(\cal M)$
\begin{equation}
\Psi^f=D^b(f)(\Psi)=e^{bf}\Psi.
\end{equation}
Note that correspondingly a testfunction $\varphi\in\cal S$ has to
transform to $\varphi^f=e^{-bf}\varphi\in{\cal S}^f$ in order to
keep $\Psi[\varphi]$ conformally invariant. Generally on a dual
space the weight should be the negative of the weight on the
original space.
In our application to the quantization of $H^f$ from (6.23), the
condition
\begin{equation}
\hat{H}^f = e^{-2f}e^{bf}\hat{H}\ e^{-bf}
\end{equation}
implies that
\begin{equation}
\hat{H}^f = -\frac{1}{2\mu}\left[ \Delta^f -\xi_c R^f\right]
+{V}^f
\end{equation}
on wavefunctions $\Psi^f = e^{bf}\Psi\in{\cal S}^{\ast f}$.
The WdW equation (6.26) is conformally equivariant
if and only if Eq. (6.26) for any $f$ is equivalent to
\begin{equation}
\hat{H}\Psi = 0
\end{equation}
where
\[
\hat{H} = \hat{H}^f\mid_{f=0}\ \mbox{and}\ \Psi =\Psi^f\mid_{f=0}
\]
are Hamilton operator and wavefunction in the harmonic time gauge.

\section{The Number $\xi_c$ in Different Dimensions}
\setcounter{equation}{0}
In Sec. 4 and 5 we have already seen that, among all possible
values for the coupling constants $\xi\equiv-a$, only
$\xi=\xi_c$ allows a conformal representation of both, the
classical and the quantum theory.
According to the examples of Sec. 4, $\xi=\xi_c$ is also a critical value,
where the relationship between the
scalarfields of two conformally related Lagrangian models,
one with minimal the other with non-minimal coupling, changes
qualitatively.

In this section we examine the dependence
of this critical number $\xi_c=\frac{D-2}{4(D-1)}$
on the dimension $D$.
Therefore we consider the prime factorization of $\xi_c$.
Table 1 lists $\xi_c=:\frac{r}{s}$ with
trivial greatest common divisor of $r$ and $s$, i.e. gcd$(r,s)=1$,
the maximal primefactor $p_{m}$ contained in either $r$ or $s$, and
the least common multiple lcm$(\xi_c):=$lcm$(r,s)$,
for dimensions $D=3\ldots 30$.
{\small
$$
\begin{array}{lccccccccccccccc}
D                &:& 3& 4& 5& 6& 7& 8& 9&10&11&12&13&14&15&16\\
\xi_c=\frac{r}{s}&:&\frac{1}{8}&\frac{1}{6}&\frac{3}{16}&\frac{1}{5}&
\frac{5}{24}&\frac{3}{14}&\frac{7}{32}&\frac{2}{9}&\frac{9}{40}&
\frac{5}{22}&\frac{11}{48}&\frac{3}{13}&\frac{13}{56}&\frac{7}{30}\\
p_{m}          &:&2&3&3&5&5&7&7&3&5&11&11&13&13&7\\
\lcm             &:&8&6&48&5&120&42&224&18&360&110&528&39&728&210\\
\end{array}
$$
$$
\begin{array}{cccccccccccccc}
17&18&19&20&21&22&23&24&25&26&27&28&29&30\\
\frac{15}{64}&\frac{4}{17}&\frac{17}{72}&\frac{9}{38}&
\frac{19}{80}&\frac{5}{21}&\frac{21}{88}&\frac{11}{46}&\frac{23}{96}&
\frac{6}{25}&\frac{25}{104}&\frac{13}{54}&\frac{27}{112}&\frac{7}{29}\\
5&17&17&19&19&7&11&23&23&5&13&13&7&29 \\
960&68&1224&342&1520&105&1848&506&2208&150&2600&702&3024&203
\end{array}
$$
\begin{center}
Table 1: $p_{m}$ and lcm of $\xi_c$ for $D=3,\ldots,30$.
\end{center}
}
We see: The smaller $\lcm(\xi_c)$, the simpler
$\xi_c$ is as a fraction.
$\lcm(\xi_c)$ has its lowest value for
$D=6$, followed by
$D=4$ and $D=3$. In these dimensions $\xi^{-1}_c$ is just an integer.
Besides $D=3,4$ and $6$, the next best is $D=10$ with $\xi_c=\frac{2}{9}$.

Note that in any dimension $D=4i+2, i\in\N$,
$\lcm(\xi_c)$ is lower than for $D-2,D-1$ and $D+1$.
When we admit for the rational composition
only the first $3$ prime numbers and  consider only dimensions $D$
in which $p_m\leq 5$,
then the lowest value of $\lcm(\xi_c)$ for $D>10$ is $D=26$.

The special simplicity of $\xi_c$ in dimensions $D=3,4,6$ and $10$
might be related to an interplay between
conformal symmetry breaking in subspaces of the universe
and the selection of its dimensions.
Resonances of fundamental frequencies coupled by $\xi_c$
might induce dimensional reduction or compactification.
Unfortunately a sufficient explanation
is presently not at hand.
\np
\noindent
\section{\bf Conclusion}
\setcounter{equation}{0}
We have emphasized that conformal coordinate transformations (1)
have to be distinguished sharply from conformal transformations of
the geometry (2), especially in Lagrangian models.
Similarily conformal equivalence transformations (2) of the
classical Lagrangian models and minisuperspace conformal transformations (3)
are conceptually very different proceedures, which have to be kept
apart very carefully. An active conformal transformation (2) of a
pure $D$-dimensional geometry yields a coordinate transformation on the
(mini-)superspace geometry. For multidimensional geometries (2.12)
conformal transformations of the factorspaces lead just to
translations in the corresponding
coordinates on minisuperspace $\cal M$.

Using invariance under (1) we have initially compared natural time gauges in
multi-dimensional universes:
(i) synchronous time,
(ii) conformal times of different factor spaces,
(iii) mean conformal time and (iv) harmonic time.
Transitions between them are given by special conformal coordinate
transformations.

As an example for invariance under (2) the conformal transformation
between the minimal coupling model
and the conformal coupling
model has been performed in arbitrary dimensions $D$.
The obtained conformal factor and scalar field
are in agreement with the result of Ref. \Xa.

By Eq. (4.26) the generalization of the scalar field
from the conformal coupling case to that of arbitrary coupling $\xi$
has been found in arbitrary dimension $D$.
For $D=4$ this expression corresponds to the earlier result of expressions
Eqs. (3.3-5) in Ref. \FM\ qualitatively.
(Note also that Eq. (5) in Ref. \Pa\ holds
only with $\xi=\xi_c=\frac{1}{6}$.)

At $\phi^2=\xi^{-1}_c$ there is a singularity of the conformal
transformation.
Hence the conformal equivalence only holds separately in the ranges
$\phi^2<\xi^{-1}_c$ and $\phi^2>\xi^{-1}_c$.

It is a characteristic feature
that natural time gauges (1) are not preserved under
conformal transformation (2)
of geometry.

Similarily we have no reason to expect that classically equivalent
conformal models (2) could by canonical quantization have
minisuperspace conformal WdW equations equivalent under (3).
This is specially evident, when the minisuperspace containes
also data beyond pure geometry, e.g. a scalar field.
While a scalar field coupled to $D$-dimensional geometry
transforms to the scalar field of the equivalent model by a  complicated
integral transform (see e.g. Eq. (4.26)),
on minisuperspace $\MS$ it is just described as an additional
coordinate, on equal footing with those from the scale factors of geometry.
Hence, from the conceptual point of view,
the attempt to treat a tensor field (the geometry) and
a scalar field on a common geometrical footing
might be questionabel in the context of canonical
(minisuperspace) quantization.
For a pure geometry however
minisuperspace can be understood better, at least for
the multidimensional geometry (2.12).

Nevertheless it remains an interesting
question for further investigations to find out, how  solutions
of the WdW equation corresponding to equivalent solutions of equivalent
Lagrangian models are related.
Interesting solutions of the WdW equation for a homogeneous
scalar field minimally coupled to multidimensional
geometry have been obtained
in Refs. \BlRZ, \Iv, \aZh\ and \bZh.

The minisuperspace $\cal M$ for a pure geometry with
nontrivial ''internal'' factor spaces $M_2,\ldots,M_n$, or $\MS$ for
geometry coupled to a scalar field,
is conformally Minkovskian.
The negative eigenvalue of its metric $G$
can be associated
with an expanding ''external'' factor space $M_1$
yielding a natural cosmological ''arrow of time'' (see Ref. \Ze).

Finally in Sec. 7 we have seen, that besides
playing a distinguished role for invariance
under both (2) and (3), the conformal coupling  $\xi_c=\frac{D-2}{4(D-1)}$
indicates number theoretically
distinguished dimensions $D=3,4,6$ and $10$, which are (besides $D=1,2$)
the most important subspace dimensions appearing in realistic models
of the universe. Further investigations will have to find
a satisfactory explanation of this coincidence.
\nl\nl
{\Large {\bf Acknowledgements}}
\nl\nl
Support by
DFG grant Bl 365/1-1
is gratefully acknowledged.
The author thanks both, the
Projektgruppe
Kosmologie at Universit\"at Potsdam, especially H.-J. Schmidt,
and the
Gravitationsprojekt with U. Bleyer, for their hospitality
and support.
Thanks for critical comments go to
U. Bleyer, H.-J. Schmidt and A. Zhuk.

\np\noindent
{\Large {\bf References}}
\nl\nl
$^{\BlRZ}$  U. Bleyer, M. Rainer and A. Zhuk, {\em Conformal
Transformation of Multidimensional Cosmological Models
and their Solutions}, Preprint FUB-HEP/94-3, FU Berlin (1994).
\nl
$^{\Bl}$  U. Bleyer, {\em Multidimensional Cosmology}, p. 101-11 in:
{The Earth and the Universe
(A Festschrift in Honour of Hans-J\"urgen Treder)},
ed.: W. Schr\"oder, Science Ed., Bremen (1993).
\nl
$^{\Iv}$  V. D. Ivashchuk, V. N. Melnikov, A. I. Zhuk, Nuovo Cim. B
{\bf 104}, 575 (1989).
\nl
$^{\bSch}$  H. J. Schmidt, {\em The Metric in the Superspace of
Riemannian Metrics and its Relation to Gravity}, p. 405 in:
Diff. Geom. and Appl., ed.: D. Krupka (Brno 1989).
\nl
$^{\Sw}$  S. T. Swift, J. Math. Phys. {\bf 33}, 3723 (1992);
J. Math. Phys. {\bf 34}, 3825 (1993);  J. Math. Phys. {\bf 34}, 3841 (1993).
\nl
$^{\GMSS}$  S. Gottl\"ober, V. M\"uller, H.-J. Schmidt
and A. A. Starobinsky, Int. J. Mod. Phys. D {\bf 1}, 257 (1992).
\nl
$^{\FM}$  T. Futamase and K. Maeda, Phys. Rev. D {\bf 39}, 399 (1989).
\nl
$^{\Pa}$  D. N. Page, J. Math. Phys. {\bf 32}, 3427 (1991).
\nl
$^{\aSch}$  H. J. Schmidt, Phys. Lett. B {\bf 214},
519 (1988).
\nl
$^{\Ha}$  J. J. Halliwell, Phys. Rev. D {\bf 38}, 2268 (1988).
\nl
$^{\Gu}$  M. C. Gutzwiller, {\em Chaos in Classical and
Quantum Mechanics}, p. 197, Series: Interdisciplinary Applied Mathematics
{\bf 1}, Springer-Verlag, New York (1988).
\nl
$^{\Ma}$  K. Maeda, Phys. Rev. D {\bf 39}, 3159 (1989).
\nl
$^{\Xa}$  B. C. Xanthapoulos and Th. E. Dialynas, J. Math. Phys. {\bf 33},
1463 (1992).
\nl
$^{\Ga}$  D. V. Gal'tsov and B. C. Xanthopoulos, J. Math. Phys. {\bf 33},
273 (1992).
\nl
$^{\Gi}$  G. W. Gibbons and S. W. Hawking, Phys. Rev. D {\bf 15}, 2752
(1977).
\nl
$^{\Ze}$  H. D. Zeh, {\em The Physical Basis of the Direction of Time},
2nd. ed. Springer-Verlag (Heidelberg, 1991).
\nl
$^{\aCh}$  T. Christodoulakis and J. Zanelli,
Nuovo Cim. B {\bf 93}, 1 (1986).
\nl
$^{\bCh}$  T. Christodoulakis and J. Zanelli, Class. Quantum Grav.
{\bf 4}, 851 (1987).
\nl
$^{\aZh}$  A. I. Zhuk, Class. Quant. Grav. {\bf 9}, 2029 (1992).
\nl
$^{\bZh}$  A. I. Zhuk, Sov J. Nucl. Phys. {\bf 55}, 149 (1992);
Phys. Rev. D {\bf 45}, 1192 (1992).
\nl

\end{document}